\documentclass[12pt]{article}
\usepackage{graphics}
\usepackage{epsf}
\usepackage{epsfig}

\def\gappeq{\mathrel{\rlap {\raise.5ex\hbox{$>$}}
{\lower.5ex\hbox{$\sim$}}}}

\def\lappeq{\mathrel{\rlap{\raise.5ex\hbox{$<$}}
{\lower.5ex\hbox{$\sim$}}}}

\def\Toprel#1\over#2{\mathrel{\mathop{#2}\limits^{#1}}}

\begin{document}
\pagestyle{empty}
\vspace*{3mm}
\begin{center}
{\Large \bf  
Reconsidered  estimates of the 10th order QED contributions 
to the muon anomaly} \\ 
\vspace{0.1cm}
{\bf A.L. Kataev}\\
\vspace{0.1cm}
Institute for Nuclear Research of the Academy of Sciences of 
Russia,\\ 117312, Moscow, Russia\\
\vspace{0.1cm}
\end{center}
\begin{center}
{\bf ABSTRACT}
\end{center}
\noindent
The problem of estimating the   
10th order QED  corrections 
to the muon anomalous magnetic moment is reconsidered.  
The  incorporation of the  recently improved   contributions 
to the $\alpha^4$ and $\alpha^5$- corrections to $a_{\mu}$ within the 
renormalization-group inspired scheme-invariant approach
leads to the estimate $a_{\mu}^{(10)}\approx 643 (\alpha/pi)^5$. 
It is in good agreement with  
the estimate  $a_{\mu}^{(10)}= 663(20) (\alpha/\pi)^5$,
 obtained by Kinoshita and Nio 
from  the   numerical calculations of  
2958 10-th order diagrams,  which are considered 
to be more  
important than the  still uncalculated 6122 10th-order  
$m_{\mu}/m_e$-dependent vertex graphs, and 
12672 5-loop diagrams, responsible for the mass-independent 
constant contribution both to $a_{\mu}$ and $a_e$.   
This confirms Kinoshita and Nio guess about dominance  
of the 10-th order diagrams calculated by them.  
Comparisons with 
other estimates of the $\alpha^5$- 
contributions to $a_{\mu}$, which exist in the literature,   
are presented.  
\vspace*{0.1cm} 
\noindent 
\\[3mm]
\vfill\eject

\setcounter{page}{1}
\pagestyle{plain}

\section{Introduction}

The problem of estimating  higher order perturbative 
corrections to the  anomalous magnetic moments was first 
formulated by Feynman. In his 1961 talk at 
one of the Conferences from the Solvey series 
he mentioned `` As a special 
challenge, is there any method of computing the 
anomalous magnetic moment of the electron which, on
the first rough approximation, gives a fair 
approximation to the $\alpha$-term and a crude 
one to $\alpha^2$ term, yielding a rough estimate 
to $\alpha^3$ and beyond ?'' \cite{Feynman:1961fj}.
This challenge was proposed when only two 
corrections  from the perturbative series for the  
anomalous magnetic 
moment of the electron $a_e$ 
\begin{equation} 
\label{ae}
a_{e}=\sum_{i\geq 1} A_i\bigg(\frac{\alpha}{\pi}\bigg)^{i} 
\end{equation}
were known. The coefficient 
\begin{equation} 
A_1=\frac{1}{2} 
\end{equation} 
was calculated   by Schwinger \cite{Schwinger:1948iu}, while 
the numerical value of the second coefficient 
\begin{equation}
A_2=-0.328478965 \dots 
\end{equation}
is  known from the independent  analytical calculations of 
Sommerfield \cite{Sommerfield}, Petermann 
\cite{Petermann}, and  Terentiev 
\cite{Terentiev}.  
An  attempt to answer the Feynman challenge was made  in Ref. 
\cite{Drell:1965hg}. Using dispersion theory the authors 
of Ref.\cite{Drell:1965hg}
managed to reproduce 90$\%$ of the $A_2$-term and presented 
the following  estimate of the 
third  coefficient 
of the expansion of $a_e$ in powers of  $(\alpha/\pi)$ :      
\begin{equation}
\label{est1} 
A_3 \approx +0.15 \dots 
\end{equation}
An analog of the  approach  of 
Ref. \cite{Drell:1965hg} was developed
in Ref. \cite{Parsons:1967sj}.
By fixating  of a cut-off in one of the basic  relations
in the dispersion theory approach,  
they were able to reproduce 
the   fourth order term   
$A_2(\alpha/\pi)^2$  approximately \cite{Parsons:1967sj} and this 
led to 
the estimate  
for the  $(\alpha/\pi)^3$ 
coefficient    
\begin{equation}
\label{est2}
A_3\approx +0.13
\end{equation} 
in agreement with the estimate of Eq. (\ref{est1}). 
Thus, initially in extracting information 
on   the value of  $(\alpha/\pi)^3$-contributions to $a_e$ 
 the dispersion technique was considered  the  proper tool  
for estimating higher-order corrections \cite{Parsons:1967sj}.
However, it was realized later on,  that this method  
cannot give information on the total value of  the 6-th order 
contributions to $a_{e}$, since the contribution 
of several sets of  three-loop diagrams 
cannot be modeled within this   
approach.  The first contribution, missed in the estimating procedures 
used in Refs. \cite{Drell:1965hg,Parsons:1967sj},  
is the 
diagram with external electron vertex and internal photon line, 
dressed by two electron loops. It  
 was analytically  calculated in Ref.    
\cite{MR}.  Another important  contribution, not taken into account  
in the estimates of Refs.\cite{Drell:1965hg,Parsons:1967sj}, 
is the subset of three-loop diagrams 
with a light-by-light-scattering subgraph. 
This contribution  was  first evaluated  numerically  in 
Ref. \cite{Aldins:1970id}, and analytically almost 
twenty years later in Ref. \cite{Laporta:1991zw}. The analytical 
calculations of all 6-th order diagrams were completed by Laporta 
and Remiddi  
in Ref. \cite{Laporta:1996mq} with the 
help of the computer symbolic manipulations 
technique. The numerical form of the results 
obtained in  Ref. \cite{Laporta:1996mq}  read :
 \begin{equation}
A_3=  +1.181241456 \dots
\end{equation} 
Thus, the estimates of 
the 6-th order contribution to $a_e$, made in  
Refs. \cite{Drell:1965hg,Parsons:1967sj},  gave the  correct  sign 
and pushed ahead  the interest 
in calculating the graphs not included in these estimates.
In its turn, the desire to perform complicated   
calculations led
to   the development 
of the symbolic manipulation computer programs 
(for a review see Ref. \cite{Veltman}), 
which played an  essential role in obtaining    
final expression for the 6-th order corrections 
to $a_{e}$ \cite{Laporta:1996mq}  in the analytical form.
Note, that the analytical calculation of the order $\alpha^3$ correction
to $a_{e}$, completed by Remiddi and Laporta  in Ref. \cite{Laporta:1996mq}, 
was continuing over 27 years after the first analytical 
calculation of the set 3-loop of  diagrams  performed
in Ref.\cite{MR}. Several theoreticians, namely     
Barbieri, Caffo, 
Levine, Remiddi, Roskies and  Laporta, contributed to the calculations of 
different diagrams during  this long-termed project
(see citations of the related works in Ref. \cite{Laporta:1996mq}).
Thus, the problem discussed at the end of  1960-s in the  Theory Division of 
INP, Gatchina  of starting from scratch   
and completing calculations of this correction rather fast \cite{Lipatov}, 
was really unsolvable problem. This was correctly understood by the members 
of this theoretical group after counting the number  and drawing 
the typical representatives of the  
3-loop diagrams, which are contributing to $a_e$.

\section{Previous status of the estimates of the 10-th order QED 
corrections to $a_{\mu}$ }

The story of the estimation of  the value  of the 10-th order 
contributions to the muon  anomalous magnetic moment $a_{\mu}$,  
developed in a  rather similar way  to the    
studies of the 
QED corrections to  $a_e$ described above. 
The first estimate, namely 
\begin{equation}
\label{est}
a_{\mu}^{(10)}\approx 570(140)\bigg(\frac{\alpha}{\pi}\bigg)^5~~~~,
\end{equation}
was given in Ref. \cite{Kinoshita:1990wp}.
It  was based on the explicit numerical  calculations of 
the diagrams from the  
set of the 10-th order contributions to $a_{\mu}$
with  the electron light-by-light-type subgraph, 
which has internal photon lines, dressed by two 
electron bubbles (an  example of these diagrams is shown 
in  Fig.1).

\begin{figure}[!ht]
\begin{center}
\includegraphics[width=0.5\columnwidth,height=0.4\columnwidth]{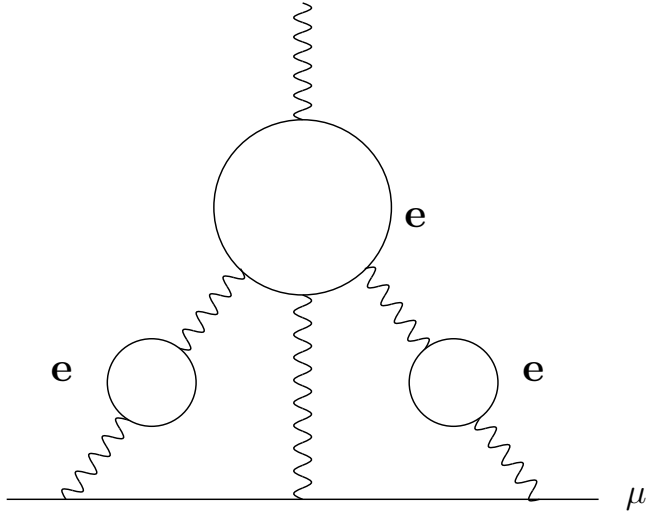}
\caption[]{The example of diagram, which was explicitly calculated 
in the process of deriving first 
estimate of the 10-th order QED  correction to $a_{\mu}$. }
\end{center}
\end{figure}

The idea to  evaluate  the contributions of these
diagrams first of all was based on the  
observation  that the most important 
contributions 
to the 6-th order and 8-th order QED corrections   
to $a_{\mu}$ came  from  light-by-light type graphs 
with an  internal electron loop insertion, coupled to the external 
muon line by three photon propagators.  
At  6-th  order 
the direct numerical   calculations of these three-loop 
diagrams 
with electron light-by-light scattering insertions  were 
calculated first in 
Ref.\cite{Aldins:1970id} and  essentially improved later on 
in  Ref. \cite{Kinoshita:1988yp}.
It is worth mentioning  here that this important 
contribution is now known   
analytically  \cite{Laporta:1992pa}. Moreover, this     
expression  was 
confirmed and even improved  
by calculating analytically  several additional terms, proportional to 
$(m_{e}/m_{\mu})^{\rm 2k}$ ($3\leq {\rm k} \leq 5$)  
in  Ref. \cite{Kuhn:2003pu}. 
At the four-loop level the dominating role of the 
``light-by-light-type'' effects were revealed  in the process 
of the numerical calculations of Ref.  \cite{Kinoshita:1990wp}. 

However, as in the case of  
the 6-th order correction to $a_{e}$, 
the estimates of the 10-th order corrections 
to $a_{\mu}$ of Eq. (\ref{est}) demonstrated 
that there are additional important contributions 
to $a_{\mu}$, which first  appear at  five-loop 
order. 
Indeed, as  was  discussed  in 
Ref.\cite{Karshenboim:1993rt} the 
subsets of diagrams  depicted in Fig.2 and Fig.3  
may also  generate sizable  
additional contributions to the 10-th order QED  corrections  
to $a_{\mu}$. 

\begin{figure}[!ht]
\begin{center}
\includegraphics[width=0.5\columnwidth,height=0.4\columnwidth]{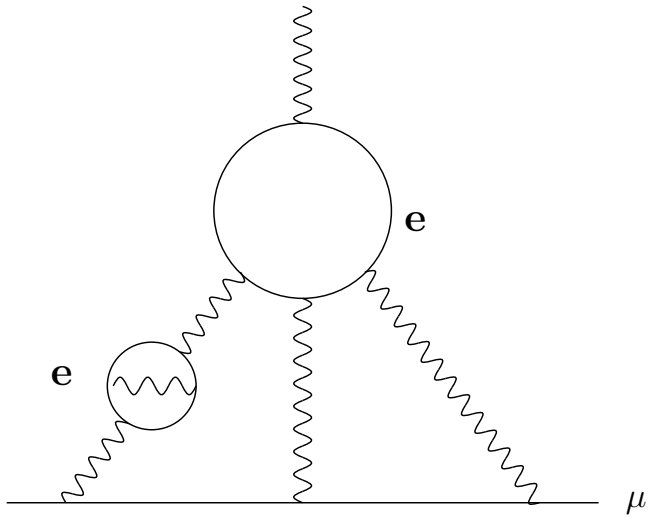}
\caption[]{The diagram, estimated first  in Ref.\cite{Karshenboim:1993rt} 
using the RG-based arguments.}
\end{center}
\end{figure}

\begin{figure}[!ht]
\begin{center}
\includegraphics[width=0.5\columnwidth,height=0.4\columnwidth]{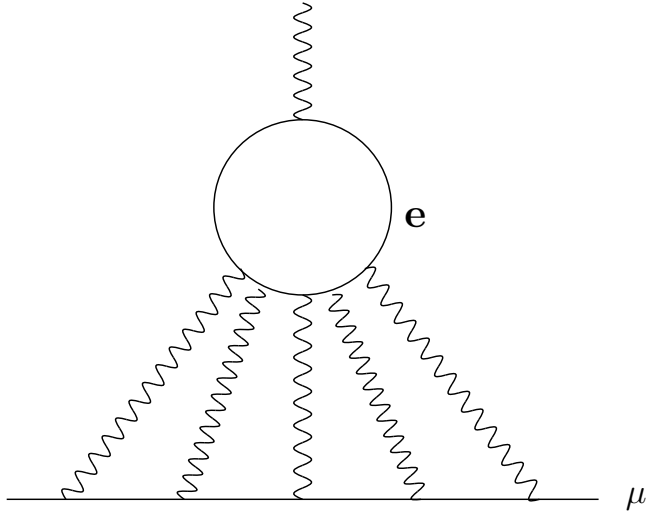}
\caption[]{The example of typical  diagram, which appear first 
at the 10-th order.}
\end{center} 
\end{figure}

The diagrams of   Fig.2 contain  
two-loop photon vacuum polarization 
insertions into one of the three internal photon lines, which connect 
electron light-by-light type subgraphs with the  external 
muon line. 
To find the approximate value 
of  this contribution  the asymptotic expression 
for the  two-loop photon vacuum polarization function was  
taken into account. As a result, the 
following estimate 
was obtained
\cite{Karshenboim:1993rt}: 
\begin{equation}
\label{vp}
a_{\mu}^{(10)}(\rm Fig.2) \approx 176(35)
\bigg(\frac{\alpha}{\pi}\bigg)^5~~~.
\end{equation}  

Another important additional contribution to  $a_{\mu}^{(10)}$ 
comes from a set  of diagrams, which appear first   
at this 5-loop level. They contain an internal electron loop, 
connected by five internal photon  lines to the external  muon line 
(see Fig.3). It was shown in Ref. \cite{Elkhovsky:1989cn},
that the expression for this
diagram has a structure  typical  of the class of 
light-by-light-type-scattering 
diagrams, namely, it contains a  
$\rm{ln}(m_{\mu}/m_e)$-term,  
which has a non-renormalization group origin. 
The general expression for this contribution  
can be written in the 
following form \cite{Elkhovsky:1989cn}:
\begin{equation}
\label{ns}
a_{\mu}^{(10)}(\rm Fig.3) \approx 
\bigg({\bf C_1} \pi^4  \rm{ln}(m_{\mu}/m_e) +{\bf C_2}\bigg)
\bigg(\frac{\alpha}{\pi}\bigg)^5~~~.
\end{equation} 
The numerical value of the coefficient $\bf{C_1}$=0.438, 
cited in Ref.\cite{Karshenboim:1993rt}, comes  
from calculations of Ref.\cite{MY}.    
Assuming, that the unknown  constant term $\bf{C_2}$  of Eq.(\ref{ns}) 
is not very large,  the author 
of Ref. \cite{Karshenboim:1993rt}
estimated this contribution   as: 
\begin{equation}
\label{nsn}
a_{\mu}^{(10)}(\rm Fig.3)=185(85)\bigg(\frac{\alpha}{\pi}\bigg)^5~~~,  
\end{equation} 

The value  for  one more set of 
the five loop light-by-light-type 
diagrams, not considered  in Ref.\cite{Kinoshita:1990wp}, and 
depicted in Fig. 4, was estimated 
in Ref.\cite{Karshenboim:1993rt} as 
\begin{equation}
\label{guess}
 a_{\mu}^{(10)}(\rm Fig.4)=\pm 40\bigg(\frac{\alpha}{\pi}\bigg)^5~~~, 
\end{equation}
without giving any theoretical arguments.

\begin{figure}[!ht]
\begin{center}
\includegraphics[width=0.5\columnwidth,height=0.4\columnwidth]{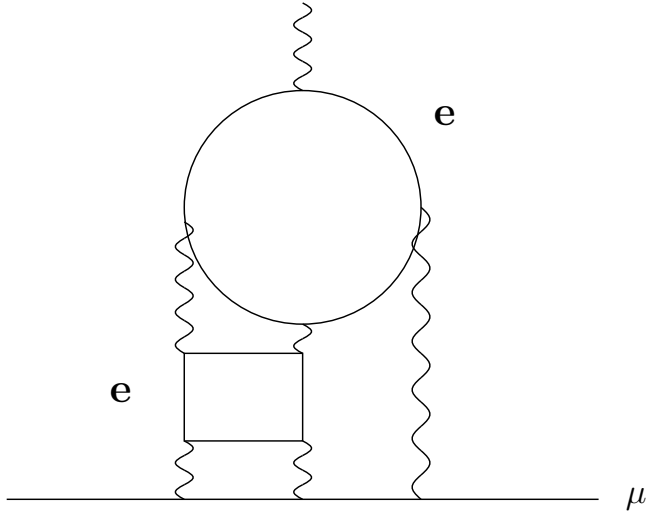}
\caption[]{The diagram, which was first considered  
in Ref.\cite{Karshenboim:1993rt} in the process 
of deriving the second  estimate of the 10-th order correction to $a_{\mu}$.}
\end{center}
\end{figure}

Combining Eq.(\ref{est}), Eq.(\ref{vp}),
Eq.(\ref{nsn}) and Eq.(\ref{guess}),  the author of 
Ref.  \cite{Karshenboim:1993rt} presented the following 
modified  estimate 
for  the  10th-order QED contributions to $a_{\mu}$
\begin{equation}
\label{SK}
 a_{\mu}^{(10)}
(\rm Ref.\cite{Karshenboim:1993rt})
= 930(170) \bigg(\frac{\alpha}{\pi}\bigg)^5~~ 
\end{equation} 
which was used in a  number of phenomenological works 
on the subject (see e.g. Ref.\cite{Czarnecki:2001pv}).

Some time later the question of the    
contribution of the   10-th order QED corrections  
to $a_{\mu}$ was  analyzed 
in Ref.  \cite{Kataev:1994rw}
using the   approaches closely 
related to the renormalization-group (RG) method, described in detail 
in Ref. \cite{Bogolyubov:1980nc},   
namely the principle of minimal sensitivity  \cite{Stevenson:1981vj}
and the effective charges (ECH) approach 
\cite{Grunberg:1980ja,Grunberg:1982fw} (for the analysis of theoretical 
problems  
and definite phenomenological applications  
of   the ECH-approach  see Refs.\cite{Krasnikov:1981rp}-
\cite{Kataev:1981gr}).

In the process of applying  the method of Ref. \cite{Kataev:1994rw}, 
tested in QCD at the $\alpha_s^3$-level in 
Ref. \cite{Kataev:1995vh},  
explicit calculations of the  
four-loop  contributions to $a_{\mu}$ \cite{Kinoshita:1990wp}
and of  the QED RG $\beta$-function in the on-shell-scheme 
\cite{Broadhurst:1992za} were used. The
expression of  Eq.(\ref{nsn})  for the new  set of diagrams of Fig.3,  
first appear in  10-th order, 
was  added to the numbers  which result   from the {\bf separate}  
application of the 
renormalization-group inspired  approach of Ref. \cite{Kataev:1994rw}  
to the set   of 
graphs with external muon vertex (Set I in the notations 
of Ref.\cite{Kataev:1994rw}, see Fig. 5),    
and to the  
set   of  diagrams which is generated by  
the three-loop light-by-light-scattering 
subgraphs  with three internal photon propagators coupled to the external 
muon line  
( Set II in the 
notations  of Ref. \cite{Kataev:1994rw}).

\begin{figure}[ht]
\begin{center}
\includegraphics[width=0.5\columnwidth,height=0.4\columnwidth]{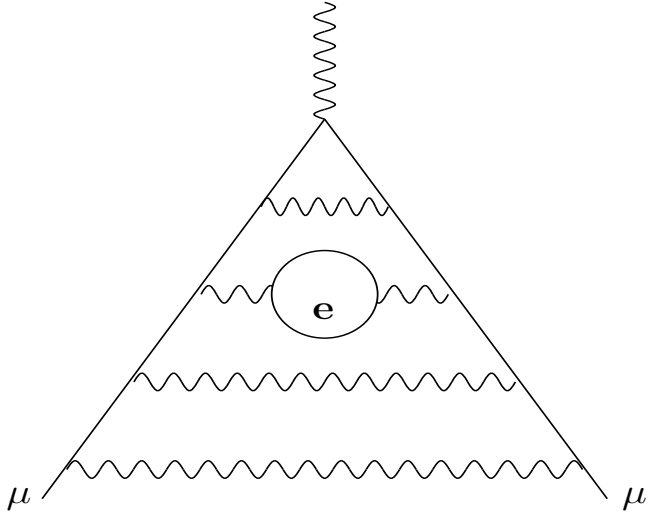}
\caption[]{The example of the diagrams which produce RG-controllable 
$\rm{ln(m_{\mu}/m_e)}$-terms in the expression for $a_{\mu}$.  }
\end{center}
\end{figure}

The estimate   
of the  10-th order contribution from the 
Set I  of diagrams  obtained in Ref.\cite{Kataev:1994rw},   is 
\begin{equation}
\label{vest}
 a_{\mu}^{(10)} ( {\rm Set~ I})\approx 
50\bigg(\frac{\alpha}{\pi}\bigg)^5~~~.
\end{equation}  

The second part of the whole 10-th order 
estimate of Ref.\cite{Kataev:1994rw},   which came  
from the consideration of the light-by-light-type graphs, 
was 
\begin{equation}
\label{lblest}
a_{\mu}^{(10)} ( {\rm Set~ II})\approx 
520 \bigg(\frac{\alpha}{\pi}\bigg)^5~~~.
\end{equation} 
It should be compared with the estimate, based on summing 
the number in Eq.(7), which comes from 
the  direct 10-th order calculations of Ref. \cite{Kinoshita:1990wp},
and  the  number of Eq.(8), which is the result 
of application of renormalization-group inspired formalism in the way, 
different from our work.One can see, that within existing huge error-bars 
this result  of the developed in Ref.\cite{Kataev:1994rw} RG-inspired 
approach (namely, Eq.(14)) is in satisfactory  agreement 
with  sum of Eqs.(7) and Eq.(8).

The total  10-th order  estimate 
of Ref.~\cite{Kataev:1994rw},  namely  
 \begin{equation}
\label{fin}
a_{\mu}^{(10)} ({\rm tot})\approx 
750 \bigg(\frac{\alpha}{\pi}\bigg)^5~~~, 
\end{equation} 
is just  the sum of 
Eq. (\ref{vest}), Eq.(\ref{lblest}) and 
Eq. (\ref{nsn}). It should be compared with the estimate, given in 
Eq. (12). We will see later on, that the improved explicit  
calculations of the 10-th order corrections 
of Ref.\cite{Kinoshita:2005sm} are in better agreement with the 
estimate of Eq. (15),
obtained in Ref.\cite{Kataev:1994rw},  than with estimate of Eq.(12), 
proposed in Ref. \cite{Kinoshita:2005sm}.  Moreover, the improved 
RG-inspired analysis will allow  us to get even the number,
which is even closer to the results of the improved calculations 
of Ref.\cite{Kinoshita:2005sm}, than Eq.(15).

\section{The improved renormalization-group inspired estimates}

\subsection{Preliminaries}

During several the  last   years   
new  results of  calculations of a number of 8-th and 10-th 
order QED contributions to $a_{\mu}$ have appeared in the 
literature (see the review of Ref.\cite{Passera:2004bj} 
and the original works of  Refs.\cite{Kinoshita:2004wi}, 
\cite{Kinoshita:2005zr} and 
\cite{Kinoshita:2005sm}).
In view of this, it is  interesting to 
study the predictive   possibilities of the scheme-invariant methods, 
incorporating 
new  4-loop and 5-loop exact  results
into the machinery of the approach of Ref.\cite{Kataev:1994rw}. 
This will allow us to understand whether the relative predictive 
success of the RG-inspired method of Refs.\cite{Kataev:1994rw},
\cite{Kataev:1995vh} is an  accident or it has a  more solid 
background.   

The total value of the 5-loop 10-th order QED correction  
to $a_{\mu}$ involves  21752 diagrams. Among them are 
12672 mass-independent graphs, contributing both to $a_{e}$
and $a_{\mu}$. They should not give huge coefficients to the 
$O(\alpha^5)$-term of the perturbative series for the electron and 
muon anomalies. Among the remaining 9080 5-loop diagrams, 2958  were 
considered to be more important than 6122 vertex graphs.  
\cite{Kinoshita:2005sm}. They contain $\rm{ln}(m_{\mu}/m_e)$ terms, 
which arise from light-by-light-scattering subgraphs and 
vacuum-polarization insertions.    
Numerical calculations, performed by Kinoshita and Nio  
 of the dominant 2958 10-th order graphs, resulted 
in the following    
more definite estimate of the 10-th order QED 
contribution to $a_{\mu}$  
\begin{equation}
\label{10new}
a_{\mu}^{(10)}({\rm Ref. \cite{Kinoshita:2005sm}})=663(20) 
\bigg(\frac{\alpha}{\pi}\bigg)^5~~.
\end{equation}
As was already mentioned above, 
it  turned out to be closer to the estimate of Ref.\cite{Kataev:1994rw}, 
than the one of Ref.\cite{Karshenboim:1993rt}
(compare Eq.(\ref{10new}) with Eq.(\ref{fin}) and Eq.(\ref{SK})).

However, since the  new 
explicit results of calculations of the 10-th order corrections to 
$a_{\mu}$ appeared in Ref. \cite{Kinoshita:2005sm}, it is the time 
to improve the knowledge about 
results of applications of the RG-inspired approach of 
Ref.\cite{Kataev:1994rw}.  
At the first stage of these studies the  
preliminary numerical expression for the set of diagrams 
of Fig.3, announced in Ref. \cite{Kinoshita:2005ti},   
\cite{clarification}
was taken into account \cite{Kataev:2005av}. 
The reduction 
of the overall value of the 
10-th order estimates 
given  in  Ref. \cite{Kataev:1994rw}
and their  striking agreement with the preliminary results 
of the direct numerical calculations, announced in 
Ref. \cite{Kinoshita:2005ti}, were 
observed \cite{Kataev:2005av}.    

It is quite understandable, that  more careful  
application of the RG 
inspired scheme-invariant approach at the  
10-th order  
necessitates the consideration of several additional 
effects. They  are:
\begin{enumerate}
\item
 the estimates of the 
 10-th order light-by-light-type ({\rm l-b-l}) contributions,  
which is  
generated by  including 
the new values of the  8-th order   
QED corrections, improved in the 
process of   calculations  of the  
Ref. \cite{Kinoshita:2004wi,Kinoshita:2005zr}. This 
improvement result in the   
following shift of the 8-th order light-by-light scattering 
terms:
\begin{equation}    
\Delta^{(8)}= a_{\mu}^{(8)}({\rm l-b-l},\cite{Kinoshita:2004wi})
-  a_{\mu}^{(8)}({\rm l-b-l},\cite{Kinoshita:1990wp})
\approx 5.2 \bigg(\frac{\alpha}{\pi}\bigg)^4~~~; 
\end{equation}
\item 
the RG  motivated 
estimates for the subset  of 10-th order graphs, which 
is generated by 8-th order graphs with  
internal light-by-light-type insertion, coupled to the external 
muon vertex by four internal photon lines (see Fig. 6). In view of the 
absence of explicit 10-th order results, this estimate was 
not considered separately in Ref. \cite{Kataev:1994rw}, but was 
included in the overall estimate of Eq.(14);    
\item 
the  explicit results for the 10-th order 
corrections from the   diagrams of Fig. 3 and Fig. 4 
published in Ref. \cite{Kinoshita:2005sm}.
\end{enumerate} 

\begin{figure}[ht]
\begin{center}
\includegraphics[width=0.5\columnwidth,height=0.4\columnwidth]{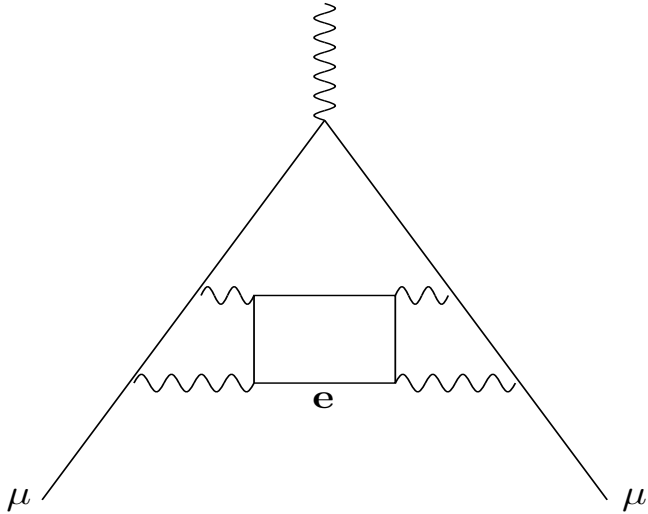}
\caption[]{The 8-th order diagram, which is generating one of the 
gauge-invariant sets of diagrams, contributing to $a_{\mu}$. }
\end{center}
\end{figure}

\subsection{Technical considerations}

In this Section we will analyze  all the  
effects discussed above.

Let us first recall the expressions for the coefficients 
of the series we will be dealing with. 

The QED contributions  to  $a_{\mu}=(g_{\mu}-2)/2$ can be   defined by 
summing 
two perturbative 
series. The first one  is the QED expression for the electron 
anomalous magnetic moment $a_e=(g_e-2)/2$, which was  
discussed in the Introduction. Rigorously speaking, 
it has the following form  
\begin{eqnarray} 
\label{ae4} 
\nonumber 
a_e &=& \sum_{i\geq 1}^{4}A_i a^{i}+O(a^2  m_e/m_{\mu})+
O(a^2  m_e/m_{\tau})
\\ 
 &=&0.5 a-0.328478965\dots a^2+1.181241456\dots a^3 -1.7283(35)a^4 
\\ \nonumber 
&+&O(a^2 m_e/m_{\mu})+ O(a^2  m_e/m_{\tau})
\end{eqnarray} 
where $a=\alpha/\pi$, and the most precise numerical  value  
for the coefficient $A_4$ was obtained in Ref. \cite{Kinoshita:2005zr}.
It  should be stressed that the available theoretical expression  
for $a_e$ agrees with  the most recent  
experimental value of $a_e$ \cite{Gabrielse:1006gg}
 at the 
level of  $10^{-10}$ (for the most recent discussions see 
Ref. \cite{Kinoshita:2005sm}). At this level 
of precision the mass dependent 
corrections, which all are $\leq 3\times 10^{-12}$  
\cite{Kinoshita:2005sm,Passera:2006gc},
can be safely neglected. 

Other, more sizable perturbative QED contributions, to $a_{\mu}$ 
can be combined 
in the series, which is  defined as 
\begin{equation}
\label{amu-ae}
a_{\mu}-a_e=A_2(m_{\mu}/m_e)+ O(m_e/m_{\mu})+ O(m_{\mu}/m_{\tau})~~.
\end{equation}
The numerical coefficients  for $A_2(m_{\mu}/m_e)$ read  
\begin{equation}
A_2(m_{\mu}/m_e)=1.0942583111(84) a^2+22.86838002(20)a^3+   
132.6823(72)a^4 +O(a^5)
\end{equation}
The $O(m_{e}/m_{\mu})$ and  $O(m_{\mu}/m_{\tau})$-corrections to 
Eq.(\ref{amu-ae}) 
are rather small (see e.g. Refs.\cite{Passera:2004bj}, 
\cite{Kinoshita:2004wi}). 
Since we are interested in estimating order $O(a^5)$-effects in the series 
 with 
increasing positive coefficients (see Eq.(20)), 
we will  neglect these  small corrections 
in all further discussions.

It should be stressed, that all diagrams 
contributing to $a_{\mu}$ at various  orders of perturbation 
theory form gauge-invariant independent sets with different 
topological structure. In view of this and in accordance   
with the  proposal 
of Ref. \cite{Brodsky:1982gc} to consider  the 
contributions to coefficients 
of the perturbative series to gauge-invariant sets  
separately 
we will divide  the expression for the  10th-order  QED contribution   
to $a_{\mu}$, defined  by five-loop diagrams,  into five  
independent gauge-invariant  contributions, namely 
 \begin{equation}
\Delta_{(10)}=\Delta_{(10)}^{(I)}+\Delta_{(10)}^{(II)}+
\Delta_{(10)}^{(III)}+\Delta_{(10)}^{(IV)}
+\Delta_{(10)}^{(V)}
\end{equation}
where Set I includes the diagram with external muon vertex 
and no light-by-light type graphs (see Fig.5), and   Set II includes 
the diagrams with light-by-light type subgraph, which has 
one external  and three internal photon propagators (this set contains 
the diagrams of the subsets of Fig.1 and Fig.2).
The generating graph  of Set III  contains a  ``box''-type 
light-by-light-scattering insertion into muon vertex, coupled 
to the external muon legs by 
four  internal photon propagators 
(see Fig. 6). 
Set IV is generated  by light-by-light scattering subgraphs
with one external and five internal muon lines 
(see Fig.3). 
Set V is generated by a  ``box''-type internal insertion
into the three-loop light-by-light-scattering  subgraph. This 
``box'' subgraph is    coupled to  
the internal photon lines  (see Fig.6). 

It is known that in the  case of the  pertrurbative series 
for $a_{\mu}$ the analytical expressions for all  high-order diagrams 
with internal electron loop insertions 
contain $\rm{ln(m_{\mu}/m_e)}$-terms.
Parts of them, namely the  ones which are coming from vacuum polarization 
 electron bubble 
insertions into internal photon lines, are governed by the RG 
method. However, the  expressions for the 
leading diagrams, which are forming sets  
with  subgraphs, generated by the effects 
of   light-by-light scattering  
on electron loops,  contain   
$\rm{ln(m_{\mu}/m_e)}$-terms which are not governed by 
the RG (typical 
examples are the diagrams of Fig.3  and  
other diagrams 
which
are generating  Set II, Set III, Set IV and Set V). 
The analytical expressions for the coefficients 
of these non-RG controllable  $\rm{ln(m_{\mu}/m_e)}$-terms
in the diagrams from Set II,  Set IV and Set V  are proportional 
to sizable  $\pi^2$-factor (see e.g. Refs. \cite{Lautrup:1977tc},
\cite{Elkhovsky:1989cn} and 
Ref. \cite{Kuraev:1989cq} 
for  the detailed discussions). Unfortunately, nothing is known 
analytically about  the coefficients of the  $\rm{ln(m_{\mu}/m_e)}$-terms
generated by the diagrams  
with the ``box'' light-by-light-scattering insertions, which 
form 
Set III and Set V.
In our further considerations we will fix all 
expressions which are  related to the appearance 
of the  diagrams which generate these 
sets  
by  
their numerical values, which are obtained  either from  
exact calculations, or from  related estimates. However, 
we will assume, that after adding  
the diagrams, which appear in higher orders of perturbation theory 
and  belong to these ``non-typical'' sets, 
the method of renormalization group will be  valid.  

Thus  we will assume, that 
the following contributions to $a_{\mu}$ 
\begin{eqnarray}
\label{aI}
a_{\mu}^{(I)}(a)&=&d_0^{(I)}a\bigg(1+d_1^{(I)}a+d_2^{(I)}a^2+d_3^{(I)}a^3 + \dots
\bigg) \\ 
\label{aII}
a_{\mu}^{(II)}(a)&=&d_0^{(II)}a^3\bigg(1+d_1^{(II)}a 
+\dots  \bigg) \\
\label{aIII}
a_{\mu}^{(III)}(a)&=&d_0^{(III)}a^4\bigg(1+ \dots \bigg) 
\end{eqnarray} 
obey the standard  RG  equations in the on-shell (OS) scheme 
\begin{equation}
(m^2\frac{\partial}{\partial m^2}+\beta(a)\frac{\partial}
{\partial a})a_{\mu}^{(I,II,III)}(a)=0
\end{equation}
with the QED RG  $\beta$-function defined as 
\begin{equation}
m^2\frac{\partial a}{\partial m^2}=\beta(a)=
\beta_0a^2(1+ c_1a+\sum_{i\geq 1}c_i a^{i+2})~~~~.
\end{equation} 
where $\beta_0=1/3$, $c_1=3/4$, $c_2=-1.26$ and 
$c_3=-1.713$ \cite{Broadhurst:1992za}. 

Note, that rigorously speaking, this 
assumption is not proved within 
the approach, developed in Ref. \cite{Lautrup:1974ic}
for the calculation of the asymptotic contributions 
to $a_{\mu}$ of the subsets 
of diagrams with 
internal photon line insertion dressed by electron loops 
inserted  
into the muon vertex (this method was 
used  for the  
 analytical evaluation of the  specific    
10-th order diagrams in Refs.\cite{Kataev:1991az,Kataev:1991cp}. 
Some of the results were  confirmed later on in Refs. 
\cite{Broadhurst:1992si,Laporta:1994md,Kinoshita:2005sm}. 
For a  recent comparison of the 
results of analytical calculations of Ref.\cite{Kataev:1991cp} and 
the numerical calculations  of Ref.\cite{Kinoshita:2005sm}, see 
Ref. \cite{Kataev:2006gx}).    

However, in  previous  studies  
of the behavior of the QED contributions 
to $a_{\mu}$ containing  light-by-light-scattering 
terms  application of the RG equations 
gave results which were not inconsistent with 
the final 10-th order estimate (see 
e.g. Eq.(14)). 
In view of this we will continue to use them in our 
improved considerations, keeping in mind our main aim: 
to   estimate the 5-loop contributions 
to $a_{\mu}^{(I,II,III)}$ using the known numerical values of the  lower-order 
coefficients $d_i^{(I)}$ ($0\leq i\leq 3)$),  $d_i^{(II)}$ 
($0\leq i\leq 1)$), $d_0^{(III)}$  and the coefficients of the QED 
$\beta$-function in the on-shell scheme. 
  
In our improved studies we will use the following numerical expressions 
for Eqs.(22)-(24)
\begin{eqnarray}
a_{\mu}^{(I)}(a)&=&0.5a+0.77a^2+3.10a^3+9.11a^4 \\
a_{\mu}^{(II)}(a)&=& 20.95a^3+126.27a^4 \\
a_{\mu}^{(III)}(a)&=&-4.43a^4 
\end{eqnarray}
where the coefficients are fixed by careful inspection of the results 
from Refs. \cite{Laporta:1991zw,Laporta:1996mq} and 
Refs.\cite{Passera:2004bj},\cite{Kinoshita:2005sm}. 
 
We will 
review  now how the  estimates procedure  developed 
in Ref. \cite{Kataev:1994rw}
may be realized  
using  the ideas of the effective charges approach of 
Ref. \cite{Grunberg:1982fw}. Within this approach one  
should nullify all higher-order corrections to physical quantities
and define for them corresponding effective $\beta$-functions.
In the case of $a_{\mu}$ we can write the  system 
of the RG-equations 
\begin{eqnarray}
a_{\mu}^{(I)}(a_{ECH}^{(I)})&=& d_0^{(I)}a_{ECH}^{(I)}(a) \\ \nonumber 
\beta_{eff}(a_{ECH}^{(I)})&=& \frac{\partial a_{ECH}^{(I)}}
{\partial a}\beta(a) \\ 
a_{\mu}^{(II)}(a_{ECH}^{(II)})&=& d_0^{(II)}a_{ECH}^{(II)~3}(a) 
\\ \nonumber 
\beta_{eff}(a_{ECH}^{(II)})&=& 
\frac{\partial a_{ECH}^{(II)}}{\partial a}\beta(a) \\  
a_{\mu}^{(III)}(a_{ECH}^{(III)})&=& d_0^{(III)}a_{ECH}^{(III)~4}(a) \\ \nonumber
\beta_{eff}(a_{ECH}^{(III)})&=& \frac{\partial a_{ECH}^{(III)}}
{\partial a}\beta(a)~~~.
\end{eqnarray}
The $\beta$-functions of the effective charges can be expressed
as 
\begin{eqnarray}
\label{ladder}
\beta_{eff}(a_{ECH}^{(I)})&=&-
\beta_0a_{ECH}^{(I)~2}\bigg(1+c_1a_{ECH}^{(I)}+\sum_{i\geq 2}
\tilde{c}_i^{(I)}a_{ECH}^{(I)~i}\bigg) \\ 
\beta_{eff}(a_{ECH}^{(II)})&=&-
\beta_0a_{ECH}^{(II)~2}\bigg(1+c_1a_{ECH}^{(II)}+\sum_{i\geq 2}
\tilde{c}_i^{(II)} a_{ECH}^{(II)~i}\bigg) \\
\beta_{eff}(a_{ECH}^{(II)})&=&-
\beta_0a_{ECH}^{(III)~2}\bigg(1+c_1a_{ECH}^{(III)}+\dots \bigg)
\end{eqnarray}
where $\tilde{c}_i^{(I)}$ and $\tilde{c}_i^{(II)}$ coefficients are 
scheme-invariant and are  defined 
using the general expressions given in   Ref. \cite{Kataev:1994rw}.

Let us now, following the considerations of 
Ref.\cite{Stevenson:1981vj,Kataev:1994rw}, re-expand 
$a_{\mu}^{(I,II,III)}(a_{ECH}^{(I,II,II)})$
in terms of the coupling 
constant $a$. In this case we will reproduce the known 
terms of Eqs.(\ref{aI})-(\ref{aIII}) and will get extra contributions,
namely
\begin{eqnarray}
a_{\mu}^{(I)}(a_{ECH}^{(I)})&=&a_{\mu}^{(I)}(a)+\Delta_N^{(I)}a^{N+1} \\
a_{\mu}^{(II)}(a_{ECH}^{(II)})&=&a_{\mu}^{(II)}(a)+\Delta_N^{(II)}a^{N+1} \\
a_{\mu}^{(III)}(a_{ECH}^{(III)})&=&a_{\mu}^{(III)}(a)+\Delta_N^{(III)}a^{N+1} 
\end{eqnarray}
which in accordance with the logic of the RG-inspired 
procedure of estimates of higher-order terms 
of Ref. \cite{Kataev:1994rw} 
will be considered as the estimates of the high  order 
contributions from these  three sets of diagrams.
Their general expressions were derived in Refs. 
\cite{Kataev:1994rw,Kataev:1995vh} and have the following form 
\begin{eqnarray}
\Delta_2^{(I)}&=&d_0^{(I)}d_1^{(I)}\bigg(c_1+d_1^{(I)}\bigg) \\
\Delta_3^{(I)}&=&
d_0^{(I)}d_1^{(I)}\bigg(c_2-\frac{1}{2}c_1d_1^{(I)}-2d_1^{(I)~2}+
3d_2^{(I)}\bigg) \\
\Delta_4^{(I)}&=&
 d_0^{(I)}d_4^{(I)}=\frac{d_0^{(I)}}{3}\bigg(3c_3d_1^{(I)}+c_2d_2^{(I)}
-4c_2d_1^{(I)~2}+2c_1d_1^{(I)}d_2^{(I)}-c_1d_3^{(I)}+
14d_1^{(I)~4} \\ \nonumber
&-&28d_1^{(I)~2}d_2^{(I)}+5d_2^{(I)~2}+12d_1^{(I)}d_3^{(I)} 
\bigg) \\
\label{lbl}
\Delta_4^{(II)}&=&
d_0^{(II)}d_2^{(II)}=\frac{2}{3}d_0^{(II)}d_1^{(II)~2}+
d_0^{(II)}d_1^{(II)}c_1 \\ 
\Delta_4^{(III)}&=&d_0^{(III)}d_1^{(III)}
\end{eqnarray}
The coefficients $d_i$ ($0\leq i\leq 4$) can be defined as  
\begin{eqnarray}
d_0^{(I)}&=&B_1 \nonumber \\
d_0^{(I)}d_1^{(I)}&=& B_2+C_2 \ln(x) \nonumber \\
d_0^{(I)}d_2^{(I)}&=& B_3+C_3 \ln(x)+D_3 \ln^2(x) \nonumber \\
d_0^{(I)}d_3^{(I)}&=& B_4+C_4 \ln(x) +D_4 \ln^2(x) + E_4 \ln^3(x)
\nonumber \\
d_0^{(I)}d_4^{(I)}&=& B_5+C_5 \ln(x) +D_5 \ln^2(x)+E_5 \ln^3(x)
+F_5 \ln^4(x)
\label{dI} \\
d_0^{(II)}&=&\overline{B}_1 \nonumber \\
d_0^{(II)}d_1^{(II)}&=&\overline{B}_2+\overline{C}_2 \ln(x) \nonumber \\
d_0^{(II)}d_2^{(II)}&=&\overline
{B}_3+\overline{C}_3 \ln(x)+\overline{D}_3 \ln^2(x) .
\label{dII}\\
d_0^{(III)}&=&\tilde{B}_1 \nonumber \\
d_0^{(III)}d_1^{(III)}&=&\tilde{B}_2+\tilde{C}_2 \ln(x) 
\label{dIII} 
\end{eqnarray}
where $\rm{ln}(x)=\rm{ln}(m_{\mu}/m_e)\approx 5.33$.
The coefficients $C_i$, $D_i$, $E_i$, $\overline{C}_i$, $\overline{D}_i$ 
and $\tilde{C}_2$ are related to the coefficients of the QED 
$\beta$-function in the OS-scheme as 
\begin{eqnarray}
C_2&=&2\beta_0B_1 \\
C_3&=&4\beta_0B_2+2\beta_1B_1  \nonumber \\
D_3&=&4\beta_0^2B_1
\end{eqnarray}
\begin{eqnarray}
C_4&=&6\beta_0B_3+4\beta_1B_2+2\beta_2B_1 \nonumber \\
D_4&=&12\beta_0^2B_2+10\beta_0\beta_1B_1  \nonumber \\
E_4&=&8\beta_0^3B_1
\end{eqnarray}
\begin{eqnarray}
C_5&=&8\beta_0B_4+6\beta_1B_3+4\beta_2B_2+2\beta_3B_1 \nonumber \\
D_5&=&24\beta_0^2B_3+28\beta_0\beta_1B_2+6\beta_1^2B_1
+12\beta_0\beta_2B_1 \nonumber \\
E_5&=&32\beta_0^3B_2+\frac{104}{3}\beta_0^2\beta_1B_1 \nonumber \\
F_5&=&16\beta_0^4B_1
\end{eqnarray}
\begin{eqnarray}
\overline{C}_2&=&6\beta_0\overline{B}_1 \nonumber \\
\overline{C}_3&=&8\beta_0\overline{B}_2+
6\beta_1\overline{B}_1 \nonumber \\
\overline{D}_3&=&24\beta_0^2\overline{B}_1 \\
\tilde{C}_2&=&8\beta_0\tilde{B}_1.
\end{eqnarray}
It is worth to stress, that Eqs.(44)-(46) reproduce all RG-controllable  
$\rm{ln(x)}$-terms, defined by Eqs.(48)-(52) and in definite cases these 
terms dominate over the contribution of the constant terms.   
In view of this,
neglecting the unknown constant term $\tilde{B}_2$, we 
can write  the RG-inspired estimate 
for the  $\Delta_4^{(III)}$-coefficient in the following form 
\begin{equation}
\Delta_4^{(III)}\approx 8d_0^{(III)}\beta_0\rm{ln \rm(m_{\mu}/m_e)}~~~.
\end{equation}

\subsection{Practical applications}

Let us consider the application of the scheme-invariant
procedure to the estimate of the 6-th and 8-th order -contribution 
to $a_{\mu}$. Using the numbers given in Eq.(27)-Eq.(29) 
\begin{equation} 
a_{\mu}^{(I)}(a)=0.5a+0.77a^2+1.76a^3+9.25a^4~~~ 
\end{equation}
where the $a^3$ and $a^4$ coefficients were estimated using Eq.(39) 
and Eq.(40) respectively. 
Note, that the  $a^3$ and $a^4$ coefficients of this series (namely 
1.76 and 9.25) should be compared  with the analogous  real physical numbers 
3.10 and 9.11 (see Eq.(27)). 

Being inspired by the  satisfactory  agreement of the 
results of applications of this approach (namely by the fact that 
it gives correct signs and reasonable values 
of the estimated $\alpha^3$ and $\alpha^4$ QED coefficients in Eq. (54))
and by its interesting outcomes in QCD   
 (see Ref.\cite{Kataev:1995vh}), we will apply  it  to 
get the improved  estimate 
the 10-th order QED contribution to $a_{\mu}$ of the first  
three terms in Eq.(21) using Eq.(41), Eq.(42) and Eq.(43) respectively. 
The results of these new estimates are:
\begin{eqnarray}
\label{one}
\Delta_{(10)}^{(I)}&=& \Delta_4^{(I)}
\approx 32 ~~,\\ \label{two}
\Delta_{(10)}^{(II)}&=& \Delta_4^{(II)}
\approx 602~~~, \\ \label{three}
\Delta_{(10)}^{(III)}&=& \Delta_4^{(III)} 
\approx -63~~.  
\end{eqnarray}
Summing them  with the results of recent direct calculations 
from Ref.\cite{Kinoshita:2005sm}
of the 
sets of the diagrams, shown  in Fig.3 and Fig.4, namely 
\begin{eqnarray}
\label{pauk}
\Delta_{(10)}^{(IV)}&=&97.123(62) \\ \label{box}
\Delta_{(10)}^{(V)}&=&-25.506(20)
\end{eqnarray}
we obtain the following estimate of the RG-inspired approach
\begin{equation}
\label{appr}
\Delta_{(10)}(\rm{est})\approx 643~~~.
\end{equation}
Note that this estimated number turned out to be in  good agreement 
with the final result of Ref.\cite{Kinoshita:2005sm}
\begin{equation}
\label{newres}
\Delta_{(10)}(\rm{calc}) = 663(20)~~.
\end{equation}
This fact allows us to make the statement that Kinoshita and Nio 
really considered   2958 graphs, which play  dominant role in forming 
 10-th order corrections to the  whole $\alpha^5$ coefficient, 
and that  still uncalculated 6122 graphs, not  
taken into account in the process of obtaining the estimate of 
Eqs.(\ref{newres}), may be not important.

\section{Comparison with other related results}

Let us first mention the sours of improvements of the order $a^5$-coefficients 
to $a_{\mu}$, estimated in Ref. \cite{Kataev:1994rw}. Note, that 
the new estimate of Eq. (\ref{one}) is slightly smaller 
than the similar old estimate of Ref.\cite{Kataev:1994rw}
$\Delta_4^{(I)}\approx50$. This effect of reduction is 
explained  by the fact that in the process of analysis 
the improved four-loop results for $a_e$, obtained in 
Ref. \cite{Kinoshita:2005zr}, were taken into account. Moreover, 
definite numerical bugs, which crept in the process of getting numerical 
number of Eq.(13),  were eliminated. 

It is also quite understandable, why the coefficient in  Eq.(14) 
differs from the similar one, obtained in this work. Indeed, this 
analog was obtained using the improved calculations of 
Ref. \cite{Kinoshita:2004wi}. 
Moreover, the presented in Eq.(14) result of Ref. \cite{Kataev:1994rw}
should be compared with the sums of Eq.(\ref{two}) and Eq.(\ref{three})
and this gives reasonable agreement of numbers (520 ${\rm vs}$ 539).

And finally, the used in this work 
value of  Eq.(\ref{pauk}), which comes 
from the  explicit calculations of  
Ref.\cite{Kinoshita:2005sm}, is smaller than 
its estimate, given in 
Ref. \cite{Karshenboim:1993rt} (see Eq.(10)). 
This difference (88) is explaining us 
the reason of the finding analogous decreasing effect in  the coefficient 
$\Delta_{(10)}(est)$ (compare Eq.(15) with Eq.(60)). 

It is also interesting to compare other discussed in this work 
approximate results  with the explicit results, which exist in 
Ref.\cite{Kinoshita:2005sm}.
First, using the 
renormalization-group formalism it is possible to derive 
the following expression for the sum of diagrams, which contain 
the diagram of Fig.2 \cite{Kataev:1994rw}:
\begin{equation}
\label{check}
a_{\mu}(\rm Fig.2)=\bigg( \bar{B}_2(\rm Fig.2)+
6\bar{B}_1\beta_1\rm{ln}(m_{\mu}/m_e)\bigg) 
\bigg(\frac{\alpha}{\pi}\bigg)^5~~ 
\end{equation}
where $\bar{B}_1=d_0^{(II)}=20.9$ is the approximate expression 
of the three-loop light-by-light scattering graph with 
electron loop and muon vertex and $\beta_1=1/4$ is the 
two-loop coefficient of the QED $\beta$-function, which is 
defined from the two-loop photon-vacuum polarization insertion.  
Thus the total value 
of the $\rm{ln}$ -contribution to Eq.(\ref{check}) 
is $6\bar{B}_1\beta_1 \rm{ln}(m_{\mu}/m_e)= 167$ 
while the  estimate of the whole 
coefficient in Eq.(\ref{check}) 
given in Ref.\cite{Karshenboim:1993rt}
is $176\pm 35$ almost coincide with this number. 
Moreover, the explicit calculations of Ref.
\cite{Kinoshita:2005sm} which gives for this contribution 
the number 181.1285(51) is supporting both the 
estimates of Ref.\cite{Kinoshita:2005sm} 
and the renormalization-group technique, used throughout this work
and leaves the room for positive contribution of the constant term    
$\bar{B}_2(\rm Fig.2)$. It is also interesting that 
 the renormalization-group inspired 
method is giving rather precise estimate for the five-loop 
diagrams, which contribute in Eq.(\ref{three}), namely the 
dressed by one electron loop contribution internal photon 
lines, which connect the electron  ``box'' 
insertion and two muon lines of the external vertex.
The RG-inspired formalism gives for this contribution the 
estimate $\Delta_{(10)}^{(III)}\approx -63$ while the explicitly 
calculated in Ref.\cite{Kinoshita:2005sm} results is 
$\Delta_{(10)}^{(III)}=-57.0633(109)$. This is one more example, when 
explicate calculations are supporting the  RG-inspired 
approach of Ref. \cite{Kataev:1994rw} used in our work. 

For the completeness of the discussions  is also worth to mention 
that some attempts to improve the precision of the truncated 
perturbative expansion were  made in   Ref.\cite{Ellis:1994qf}
by means of  the Pade resummation technique. However, it is  
difficult    to compare their estimates with the results 
discussed in this paper. Indeed, the analysis 
of Ref.\cite{Ellis:1994qf} does not allow   
to separate the renormalization-group controllable contributions 
from the ones, which are coming from the diagrams with 
the light-by-light-type structure.     

To conclude, we see, that the renormalization-group 
inspired  scheme-invariant approach is working quite  satisfactory  
for the estimates of 10-th order corrections 
to the muon anomalous magnetic moment. The estimates obtained 
 are supporting 
the applications of the results of recent calculations 
of Ref. \cite{Kinoshita:2005sm} in the phenomenological analysis.
The current difference between phenomenological and 
experimental results for $a_{\mu}$ 
(see e.g. Ref.\cite{Bennett:2006fi}), which is more straightforwardly  
related to the values of the  effects of strong interactions,  
can not be described by 
including tenth order  QED effects  in the theoretical predictions.
Indeed, they are smaller than the  experimental uncertainties.
Thus the  present value of  the order $a^5$ QED contribution to 
$a_{\mu}$ is over 0.01 times smaller than the previous order $a^4$-one,
which in tern is over 0.013 times  smaller than the  order $a^3$-contribution.
The numerical values of these terms, namely $0.4\times10^{-10}$ 
and $38.1\times 10^{-10}$ \cite{Eidelman}   
may serve the aim for fixing the errors 
of possible future $(g_{\mu}-2)$ measurements, since most recent 
one has statistical and systematical uncertainties of over 
$(\pm 5.4 \pm 3.3)\times 10^{-10}$ \cite{Bennett:2006fi}, which are higher  
that the order $\alpha^5$ QED correction.

\section{Acknowledgments}
I would like to thank T. Kinoshita for 
renewing my interest to the considerations  of the  10-th 
order QED corrections  to $a_{\mu}$, for 
inviting me to 
present the results of these studies at the  Cornell University 
and for useful discussions. I am grateful to S.L. Adler for 
his invitation to visit Institute for Advanced Study in Princeton, 
where this work was started.  
I  want   to wish  the success  to   V.V Starshenko 
in his new work at IPG Photonic Corporation.  The collaboration with him  
at the previous stage of the analogous studies was really 
pleasant. 
It is really worth  to acknowledge   
the  encouraging atmosphere of  
Theoretical Division of INR.
This work is supported  by  RFBR Grant N 05-01-00992.

\end{document}